\documentclass[twocolumn,superscriptaddress,pra]{revtex4}
\usepackage{amsmath,mathrsfs,amsbsy,amsfonts,color,graphicx,bm}

\newcommand{\bra}[1]{{\left\langle #1 \right|}}
\newcommand{\ket}[1]{{\left| #1 \right\rangle}}
\newcommand{\red}{\color[rgb]{0.8,0,0}}

\begin{document}




\title{Limitations to sharing entanglement}

\author{Jeong San Kim$^{a, b \ast}$
,\thanks{$^\ast$Corresponding author. Email: jeongsan.kim@gmail.com
\vspace{6pt}}
Gilad Gour$^{a, c}$ and~Barry C. Sanders$^\text{a}$\\\vspace{6pt} $^\text{a}${\em{Institute for
Quantum Information Science, University of Calgary, Alberta, Canada T2N 1N4}}\\
$^{b}${\em{Department of Mathematics, University of Suwon, Kyungki-do 445-743, Korea
}}\\
$^{c}${\em{Department of Mathematics and Statistics, University of Calgary, Alberta, Canada T2N 1N4}}\\
\vspace{6pt}
}
\begin{abstract}
    We discuss limitations to sharing entanglement known as monogamy of entanglement.
    Our pedagogical approach commences with simple examples of limited entanglement sharing for
    pure three-qubit states and progresses to the more general case of mixed-state monogamy relations with multiple qudits.
\bigskip
\end{abstract}

\maketitle
\section{Introduction}
\label{sec:introduction}

Entanglement is one of the most important feature of quantum
mechanics and arises from applying the quantum superposition
principle to multiple systems~\cite{PV98, HHHH09}. In its bipartite
form, entanglement is at the heart of Bell inequality violations,
which yield spacelike-separated detector correlations that exceed
the degree of correlation permitted within the framework of local
realism. Recently entanglement has been recognized as a key
consumable resource in quantum communication, for example to
teleport~\cite{BBC+93} or to perform dense coding~\cite{BW92}.
Entangled states can be used by two parties to perform secure
quantum key distribution~\cite{Eke91}, and, although entangled
states are not required for the Bennett-Brassard 1984 quantum key
distribution protocol known as BB84~\cite{BB84}, an entangled-state
analysis is used to prove the security of privacy amplification for
overcoming errors in establishing the key. Entanglement is essential
in certain models of quantum computing such as non-deterministic
gate teleportation~\cite{NC97, HZB06, MJDF08} in cases where a
deterministic gate is not feasible, and a large multi-partite state
is required as the initial quantum computing ``substrate'' for
one-way quantum computing~\cite{RB01}.

Entanglement is a shareable resource meaning that parties in a
network can transfer entanglement between themselves. These players
can share entanglement by sending shares (e.g.\ particles) through
quantum channels to each other or, alternatively, consume prior
shared entanglement and employ classical communication to
redistribute entanglement to other parties. Our concern here is to
show that severe restrictions apply to the sharing of entanglement.
These restrictions are known as monogamy relations. The terminology
refers to the concept of fidelity in marriage: if two adults are
committed to each other, then a third adult has no amorous access to
the couple. For the case of two-level systems (quantum binary
digits, or `qubits' in quantum information parlance), two maximally
entangled qubits share no entanglement or even classical
correlations whatsoever with any other qubits (this phenomenon holds
for higher level systems as well).

Monogamy of entanglement~(MoE) was first proposed by Coffman, Kundu
and Wootters in 2000~\cite{CKW00}. They conjectured MoE for a
network of many qubits but only succeeded in proving the
quantitative monogamy relation for three qubits. The conjecture
stubbornly remained unproved for several years until
the proof by Osborne and Verstraete
in 2006, which was a calculational tour de force~\cite{OV06}.
These results show that, in a quantum network comprising $n$ parties
each possessing one qubit, more entanglement shared between two
parties necessarily implies less entanglement that can be shared
between either of these parties with any other parties in the
network. Furthermore, shared entanglement between two parties even
limits the amount of \emph{classical} correlation that can be shared
with the other parties~\cite{KW04}.

Without delving into the details of entanglement measures yet, the concept of
a \emph{monogamous measure of entanglement} can be described as follows.
Let $E_{\text{A}|\text{B}}$ denote the shared entanglement between~A and~B.
That is to say $E_{\text{A}|\text{B}}$ measures the degree of entanglement between~A and~B,
but of course there is more than one choice for this measure, which concerns us later.
As MoE is concerned with more than two parties in a quantum network,
we also consider the entanglement $E_{\text{A}|\text{C}}$ shared between~A and~C
and the entanglement $E_{\text{A}|\text{BC}}$ shared between~A and the composite system comprising~B and~C.
In this notation, $E$ is monogamous if
\begin{equation}\label{monog}
E_{\text{A}|\text{BC}}\geq E_{\text{A}|\text{B}}+E_{\text{A}|\text{C}}
\end{equation}
provided that adding $E_{\cdot|\cdot}$ is meaningful.
This inequality conveys the MoE principle that the amount of entanglement shared between A and~B restricts the possible
amount of entanglement between A and C so that their sum
does not exceed the total bipartite entanglement between A and the composite BC system.

In the seminal Coffman-Kundu-Wootters paper~\cite{CKW00}, the
entanglement sharing limit of Eq.~(\ref{monog}) was proved only for
systems consisting of qubits, for which the entanglement measure~$E$
can be taken to be the square of Wootters's
concurrence~\cite{Woo98}. This squared-concurrence quantity is a
measure of entanglement known also as the ``tangle''
whose formal definition is discussed later.
The MoE relation of Eq.~(\ref{monog}) when expressed in terms of the qubit
tangle is elegant yet unsatisfying. The elegance of the expression
is evident in its simplicity and in the presence of a saturable
bound on the amount of entanglement that can be shared. The
unsatisfying aspect of this quantum-network entanglement-sharing
relation is its poor capability for direct generalization to higher
dimensions. For example, suppose that the parties hold quantum
$d$-ary digits, or `qudits', rather than qubits. Generalizing the
Wootter's tangle from the case of qubits to the case of qudits leads
to measures of entanglement that are not additive (nor
superadditive).  As any measure of entanglement that satisfies the
monogamy relation~(\ref{monog})
{\red{must be additive or super-additive}}
(see Sec.~\ref{subsec:squashed} for details) the
generalization of the qubit monogamy relation in~\cite{CKW00} to
qudits is impossible.

A natural question therefore arises: is there a measure of
entanglement satisfying Eq.~(\ref{monog}) for all dimensions?
Whereas most known measures of entanglement (like entanglement of
formation~\cite{Woo98} or the relative entropy of
entanglement~\cite{VPRK97, VP98}) are not monogamous, there exists
one measure, called the \emph{squashed entanglement}, which
satisfies Eq.~(\ref{monog}) in all dimensions. The existence of such
a measure indicates that MoE can be quantified, but since the
squashed entanglement is extremely hard to compute even numerically
(see Sec.~\ref{subsec:squashed} for more details), our knowledge
about the sharability of entanglement in general non-qubit quantum
networks is very limited.

As MoE is an interesting fundamental property of quantum mechanics
and yet is not mathematically settled for general quantum networks,
our aim here is to introduce and explain MoE in a way that avoids
the daunting mathematical complexity of the subject and instead
builds a conceptual foundation. We commence by focusing on extremal
cases of pure-state three-qubit quantum networks and progress to the
general three-qubit network. Subsequent to building this foundation,
we discuss multi-qubit quantum networks in a gentle way.
Using these concepts we show how to bound the tangle using a dual relation to MoE, sometimes called
polygamy of entanglement~\cite{GBS07, BGK09, Kim09}.

\section{Entanglement in a network and the concept of monogamy of entanglement}
\label{sec:entanglement}

Entanglement refers to quantum correlations between two or more parties.
The state shared between two parties A and~B is designated $\rho_\text{AB}$ and called a \emph{bipartite} state.
Mathematically, it is a unit-trace bounded operator on the tensor-product Hilbert space
$$\mathscr{H}_\text{A}\otimes\mathscr{H}_\text{B}$$
for $\mathscr{H}_\text{A}$ the Hilbert space for~A and~$\mathscr{H}_\text{B}$ the Hilbert space for~B.
A state is `pure' if and only if~$\rho=\rho^2$; otherwise the state is `mixed'.
A pure state can be expressed as a projector~$|\psi\rangle\langle\psi|$ in Dirac notation
with $|\psi\rangle\in\mathscr{H}$ so here we often refer to pure states as elements of Hilbert space~$\mathscr{H}$
whereas mixed states are always elements of~$\mathcal{B}(\mathscr{H})$,
i.e.\ \ bounded operators (or density matrices) acting on~$\mathscr{H}$.

 Entanglement is a key ingredient of many counter-intuitive quantum phenomena.
 Besides being of interest from a fundamental point of view,
entanglement has been identified as a non-local resource for quantum
information tasks. In particular, shared bipartite entanglement is a
crucial resource for many quantum communication tasks such as
teleportation.

Such a scenario, in which parties sharing a composite quantum system
are only able to perform local operations on their share of the
system, arises naturally (as systems~A and~B can be located far from
each other) and is quite common in quantum information theory.
Communicating classical information is considered to be easy. Local
quantum operations assisted by classical communication is known as
local operations with classical communication (LOCC), and we will
use this acronym frequently.

In quantum information science, entanglement is quantified by its
ability to overcome the LOCC restriction. Therefore, a bipartite
state is entangled if and only if it cannot be prepared by LOCC
between two parties~A and~B. This definition of an entangled state
is consistent with our intuition that the term entanglement refers
to non-local quantum correlations. That is, classical communication
cannot generate quantum correlations, and local operations cannot
generate non-local correlations. Therefore, LOCC cannot generate
entanglement nor can it increase entanglement. For this reason
entanglement must be quantified by functions that do not increase
under LOCC.

States that can be prepared by LOCC (i.e.\ , states with zero
entanglement) are called \emph{separable} states and they have a
particular form, as we now discuss. Consider two parties usually
referred to  A and B who are located far from each other, in the
sense that the LOCC restriction is applied to the composite physical
system they share. We ask what kind of states representing their
physical system they can prepare.

Clearly A can prepare her (local) physical system in some state
$\sigma_{\text{A}}$, and B can prepare his system in a state
$\sigma_{\text{B}}$. In this situation, the joint composite physical
system of A and B is represented by one state
$\sigma_\text{A}\otimes\sigma_\text{B}$. Such a state is called a
\emph{product} state and it can definitely be prepared by LOCC. In
fact to prepare a product state there is no need for classical
communication.

Consider now a situation in which A and B can prepare their physical
systems in one out of several states $\{\sigma^{(\ell)}_\text{A}\}$
and $\{\sigma^{(\ell)}_\text{B}\}$, respectively. Here the integer
$\ell$ distinguishes between the different states. Consider also a
protocol in which A choose to prepare the state
$\sigma^{(\ell)}_\text{A}$ according to a fixed probability
distribution $p_\ell$. Then, once she choose $\ell$ she communicates
it (classically, for example by telephoning) to B who then prepares
his system in the state $\sigma^{(\ell)}_\text{B}$.

At the end of this protocol the composite system of A and B is
described by the product state
$\rho^{(\ell)}_\text{A}\otimes\rho^{(\ell)}_\text{B}$ which occurs
with probability $p_\ell$. If A and B `forget' (or lose) the
information about $\ell$, then their composite physical system is
represented by a separable state $\rho_{\text{AB}}$ which is given
by
\begin{equation}
\label{eq:SEP}
    \rho_\text{AB}\equiv\sum_\ell p_\ell\;\sigma^{(\ell)}_\text{A}\otimes\sigma^{(\ell)}_\text{B}.
\end{equation}
Any bipartite state $\rho_\text{AB}$ that has the form above is called a `separable state'.

Separable states represent physical systems with no entanglement because they can be prepared by LOCC.
A bipartite quantum state $\rho_\text{AB}$ that is not separable is called an entangled state.
Recall that quantum states such as $\rho_\text{AB}$ above are square Hermitian matrices with non-negative
eigenvalues that sum to one. It is therefore not straightforward to determine whether a bipartite state
$\rho_\text{AB}$ can be written in the form~\eqref{eq:SEP} or not. Indeed,
determining whether or not a state is entangled, and how much
entanglement is inherent within the state, are hard tasks.

One reason for the difficulty is that the state~$\rho$ needs to be
known in order to evaluate entanglement, but the matrix
representation has size $d^2$ for~$d$ the dimension of the Hilbert
space. As~$d$ is exponential in the number of qubits or qudits
encoding the quantum information, tomography for determining the
full state is generically hard. In fact even with complete knowledge
of the state, deciding whether a state is entangled or not is
computationally hard.
An example of a hard problem class is NP (nondeterministic polynomial),
which refers to the class of computational problems whose best-known algorithms
require exponential time to solve but candidate solutions are easy
(polynomial time) to verify. Deciding whether a state is entangled or not
is NP-hard~\cite{Gur04, Gha10}, meaning that the problem is known to be at least as
hard as other problems believed to be in the NP class.

We do not consider this problem here, but one way that partially
overcomes this problem is to introduce entanglement
witnesses~\cite{HHH_PLA96}. An entanglement witness is an Hermitian
operator or an observable of a geometric nature (in the terminology of convex sets is called a
supporting hyperplane~\cite{PR02}) that distinguishes entangled
states from separable ones. More precisely, for every entangled
state~$\rho$ there exists an Hermitian operator $W$ (the
entanglement witness) such that the expectation value of~$W$ with
respect to~$\rho$ is negative, i.e.\ ,
\begin{equation}
\label{eq:W}
    \text{Tr}(\rho W)<0,
\end{equation}
whereas the expectation value is positive  with respect to all separable states.

One problem with entanglement witnesses is that they tend to
deliver only one-sided answers to the decision problem: if the
expectation value is negative then the state is entangled, but if
the expectation value is non-negative, whether that state
is separable or entangled is not known for sure.
Therefore, entanglement witnesses can be very useful in determining whether a
state is entangled or separable, as long as their number is kept
small.

As can be seen in Eq.~(\ref{monog}),
monogamy of entanglement is defined relative to some measure of entanglement $E$.
Measures of entanglement are non-negative real valued functions on the space of density matrices
(i.e.\ , quantum states).
They are expected to satisfy certain conditions, such as having the value zero for separable states. Perhaps the most important property of measures of entanglement
is their monotonicity, whereby different parties cannot increase the total average entanglement by LOCC.

Over the past decade many measures of entanglement have been introduced pertaining to entanglement theory.
Many of these measures are unfortunately not additive (nor superadditive), and as we see in
Sec.~\ref{subsec:squashed},
additivity does play a crucial role in MoE relations.

In fact there are only a few known entanglement measures that are
monogamous; that is, satisfying the monogamy relation~\eqref{monog}.
The first entanglement monogamy relation was introduced by Wootters
using tangle; however
it only measures the
entanglement of (possibly mixed) two-qubit states~\cite{Woo98}.
Therefore, in order to discuss monogamy of entanglement for higher
dimensional systems such as qutrits, we need to employ other
measures of entanglement later in this paper.

Whereas monogamy of entanglement is a property of multi-party
quantum correlations, here focus mainly on quantifying
\emph{bipartite} entanglement.
The main reason for this focus is
that, although Inequality~(\ref{monog}) consists of entanglement between two
parts of the composite system (note that even $E_{A|BC}$ refers to
entanglement between \emph{two} subsystems, namely, subsystem $A$
and the composite subsystem $BC$), Inequality~\eqref{monog} also yields
\begin{equation}
\label{eq:E_ABC}
    E_{ABC}\equiv E_{\text{A}|\text{BC}}- E_{\text{A}|\text{B}}-E_{\text{A}|\text{C}},
\end{equation}
which is not negative and can be interpreted (for $E$ Wootters's
concurrence) as genuine tripartite entanglement~\cite{CKW00}.
Therefore, quantifying genuine tripartite entanglement
can be obtained by quantifying bipartite entanglement among subsystems.
In this interpretation the entanglement between $A$ and $BC$ comprises
the entanglement between A and B, the entanglement between A and C,
and the tripartite entanglement among A, B, and C.

Let us now consider a tripartite quantum network with each party
holding a qubit. If a qubit is in a pure state, we can represent it
with a state $\alpha|0\rangle+\beta|1\rangle$ where $\alpha$ and
$\beta$ are complex numbers and the two-dimensional vectors
$|0\rangle,\;|1\rangle\in\mathbb{C}^2$ represents the two degrees of
freedom such as spin up and spin down of a spin-$\tfrac{1}{2}$
particle.

With this notation, if the overall three-qubit state is pure, we can write the most general shared quantum state
as a normalized vector in
$\mathbb{C}^2\otimes \mathbb{C}^2\otimes\mathbb{C}^2$:
\begin{equation}
\label{eq:3qubitpurestate}
    \left|\bm{\psi}\right\rangle_\text{ABC}
        =\sum_{\epsilon_1,\epsilon_2,\epsilon_3=0}^1\psi_{\epsilon_1,\epsilon_2,\epsilon_3}\left|
            \epsilon_1,\epsilon_2,\epsilon_3\right\rangle,
\end{equation}
where $\{\left|\epsilon_1,\epsilon_2,\epsilon_3\right\rangle\equiv
|\epsilon_1\rangle\otimes|\epsilon_2\rangle\otimes|\epsilon_3\rangle
\}$ is the so-called computational basis for three qubits. For
example, the state $|000\rangle$ represents three up spins,
$|100\rangle$ the first spin being down and remaining two spins up,
and so on. The eight coefficients
$\{\psi_{\epsilon_1,\epsilon_2,\epsilon_3}\}$ are complex numbers
subject to the normalization constraint
$\langle\bm{\psi}|\bm{\psi}\rangle=1$. Thus, the state can be
expressed as a complex vector~$\bm{\psi}$ in the projective
eight-dimensional complex space $P(\mathbb{C}^2)^{\otimes
3}=P\mathbb{C}^8$.

Suppose now that the first two qubits A and~B, share a maximally entangled state
\begin{equation}
\label{eq:singlet}
    \ket{\Psi^\pm}_\text{AB}=|01\rangle\pm|10\rangle
\end{equation}
(known as the `singlet state' for the~$-$ case),
and supress the state-normalization coefficient if it is clear by context.
The dropped normalization coefficient in Eq.~(\ref{eq:singlet}) is~$\frac{1}{\sqrt{2}}$.
Physically this state can be realized with two spin-$\frac{1}{2}$ particles, such as an electron,
where each of the two parties holds a particle in the opposite spin state,
and the state is antisymmetric under exchange of particles.

If~A and~B share a singlet, then the pure state for the composite A, B and C system must have~C's state in a simple
tensor product with the AB state. If we write C's general single-qubit state as~$|\varphi\rangle\in\mathbb{C}^2$,
then tripartite state $\left|\bm{\psi}\right\rangle_\text{ABC}$ in Eq.~\eqref{eq:3qubitpurestate} must have the form
$\ket{\Psi^-}_\text{AB}|\varphi\rangle_\text{C}$.
Thus, A and~B sharing a singlet implies that C's state is entirely independent. This is the essence of MoE.
The question is then why  C's state must be entirely independent and why qubit~C cannot also be maximally entangled with~A.
The answer to this question emerges from the no-cloning theorem.

\begin{figure}
\includegraphics[width=8cm]{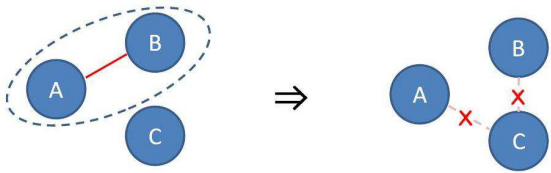}
\caption{
    Two-qubit singlet state and monogamy of three-qubit systems:
    if subsystems~A and~B are in singlet state, there cannot be any entanglement
    shared between~A and~C nor~B and~C.}
\label{fig:nosharedentanglement}
\end{figure}

The no-cloning theorem is a result of quantum mechanics that forbids the creation of identical copies of an arbitrary unknown quantum state.
As~A and~B share a maximally entangled two-qubit state,
A and~B have the requisite quantum resource to teleport an unknown quantum state from one to the other.
As shown in Fig.~\ref{fig:nosharedentanglement},
suppose that~A and~C also share a maximally entangled two-qubit state.
Then A can teleport an unknown quantum state to~C.

This set-up can be exploited to clone an unknown quantum state as follows.
A teleports the state to~B \emph{and} to~C.
Thus, this tripartite network has succeeded in copying the state:
B and~C each hold a copy now.
However, this operation violates the no cloning theorem,
which is in turn a direct consequence of the linearity of quantum mechanics~\cite{WZ82}.
If~A and~B share a maximally entangled state,
even if one of the two parties shares any entanglement whatsoever with the third party~C,
the no cloning theorem is violated.
Therefore, MoE has its foundation in the linearity of quantum mechanics.

At this point we have not discussed whether Inequality~\eqref{monog} can saturate,
and, if so, which states saturate the inequality.
To address this issue, we explore genuine tripartite entanglement among all three qubits
in a tripartite quantum network.
Unlike the bipartite case,
three qubits or more possess more than one type of entanglement.
For three qubits, two quite distinct types of
entanglement exist~\cite{WVC00} under stochastic LOCC (SLOCC)~\cite{WVC00},
I.e., random or non-deterministic LOCC whereby only desired outputs under LOCC are retained):
the W type and the Greenberger-Horne-Zeilinger (GHZ) type.
The two SLOCC classes of states are defined in terms of the W state
\begin{equation}
\label{eq:W}
    \ket{\text W}=\ket{001}+\ket{010}+\ket{100}
\end{equation}
and the GHZ state~\cite{GHZ89}
\begin{equation}
\label{eq:GHZ}
    \ket{\text{GHZ}}=\ket{000}+\ket{111}
\end{equation}
with obvious normalization coefficients suppressed.
W-class states and GHZ-class states are determined by whether a
given three-qubit state~$\left|\bm{\psi}\right\rangle_\text{ABC}$ can be converted to a W state or to a
GHZ state, respectively, by SLOCC.
Every three-qubit entangled pure state belongs to either the GHZ class or to the W class under SLOCC.

Here SLOCC is considered rather than LOCC to classify inequivalent classes
of entangled states. 
For single copies, two pure states $\ket{\psi}$ and $\ket{\phi}$ can be obtained
with certainty from each other by means of LOCC if and only if they are related by local unitaries. However, even in the simplest bipartite systems, $\ket{\psi}$ and $\ket{\phi}$ are
typically not related by local unitaries, and continuous parameters
are needed to label all equivalence classes. That is, one has to deal with infinitely many kinds of entanglement.
In this context an alternative, simpler classification would be advisable, and
one such classification is possible if we just demand
that the conversion of the states is through LOCC but without imposing that it has
to be achieved with certainty, that is SLOCC. In that case we can establish
an equivalence relation stating that two states $\ket{\psi}$ and $\ket{\phi}$ are equivalent if the parties have a non-vanishing
probability of success when trying to convert $\ket{\psi}$ into $\ket{\phi}$,
and also $\ket{\phi}$ into $\ket{\psi}$.

The GHZ class is dense in the Hilbert space of three qubits, which implies that the~$W$ class is of measure zero.
Therefore, if a state is picked randomly (say with the Haar measure) from the space of three qubits,
with unit probability it belongs to the GHZ class.
The GHZ state is special in that the state is maximally entangled with respect to every bipartition
(A$|$BC, AB$|$C, AC$|$B).
The W state is special in that it is the only state (up to local unitary operations) that saturates the monogamy inequality~(\ref{monog}), and also
it is the only state that maximizes the average bipartite entanglement as we discuss now.

Consider the reduced density matrix for AB of the~$W$ state
\begin{align}
\label{eq:Wrhoab}
    \rho_\text{AB}=&\text{tr}_\text{C}\ket{\text W}_{\text{ABC}}\bra{\text W}
        \nonumber   \\
    =&\frac{1}{3}\ket{00}_\text{AB}\bra{00}+\frac{2}{3}\ket{\Psi^+}_\text{AB}\bra{\Psi^+},
\end{align}
which is obtained by taking the partial trace over~C.
Because of permutation symmetry, the states~$\rho_\text{BC}$ and~$\rho_\text{AC}$ are similar.
The normalization of the maximally entangled state~$\ket{\Psi^+}$ is needed here to ensure that the overall
state~(\ref{eq:Wrhoab}) has unit trace and appropriate weighting in the sum.
Regardless of the chosen entanglement measure~$E$,
it should be maximized for the state~$\ket{\Psi^+}$,
and, by convention, $E=1$ for maximally entangled states and~$E=0$ for unentangled states.
Hence, the average entanglement of the decomposition of~$\rho_\text{AB}$ in~(\ref{eq:Wrhoab}) is given by
\begin{align}
\label{eq:rhoabave}
    E\left(\rho_\text{AB}\right)
        =&\frac{1}{3}E\left(\ket{00}_\text{AB}\bra{00}\right)
            +\frac{2}{3}E\left(\ket{\Psi^+}_\text{AB}\bra{\Psi^+}\right)
                \nonumber   \\
        = \frac{2}{3}.
\end{align}
Note that there are other ensemble decompositions for $\rho_\text{AB}$ than the one given in~(\ref{eq:Wrhoab}),
but it can be shown~\cite{WVC00} that decomposition~(\ref{eq:Wrhoab}) has the minimum average entanglement.
For this reason (see Sec.~\ref{sec:bipartite}), we assume $E\left(\rho_\text{AB}\right)=2/3$.
From permutation symmetry of the $W$ state,
we know that  $E\left(\rho_\text{AC}\right)=E\left(\rho_\text{BC}\right)=2/3$. Thus, the
the overall average bipartite entanglement of the three-qubit W state is given by
\begin{equation}
\label{eq:exp}
    \frac{1}{3}\left[E\left(\rho_\text{AB}\right)+E\left(\rho_\text{BC}\right)+E\left(\rho_\text{AC}\right)\right]=\frac{2}{3},
\end{equation}
which is the maximum possible value for \emph{any} three qubit mixed state~$\rho_\text{ABC}$~\cite{WVC00}.

The GHZ state has zero entanglement in this sense: each pair of parties has no shared entanglement after tracing out the third party.
Thus, the GHZ and W states are quite different types of maximally entangled three-qubit states.
The GHZ state maximizes the expected pure bipartite
entanglement between any one qubit and the other two qubits, and the W state
maximizes the expected mixed two-qubit bipartite entanglement after tracing out the third qubit.

Whereas three-qubit states nicely partition into the two non-overlapping W and GHZ classes, for four or more qubits systems
there are uncountably many SLOCC classes, thereby making the task of classifying entanglement for more
than three qubits much harder~\cite{WVC00,GW10}.
Further complicating matters, the existence of a state like the three-qubit GHZ state, with the property of being maximally entangled under every bipartition
is unique to the three-qubits case. Although there do exist states with maximal entanglement under every bipartition
for five and for six qubits, there are no such states for four qubits or more than seven qubits~\cite{GW10}, and the seven-qubit case remains an open question.

For the general multipartite network with a mixed state,
the notion of `sharability' is important in studying MoE.
A bipartite state~$\rho_\text{AB}$, which is shared by two parties~A and~B,
is called $n$-shareable if there exists an $(n+1)$-partite quantum state
$\rho_{\text{A}\textbf{B}^{(n)}}$
which is shared by one party~A and~$n$ parties \textbf{B}=\{B$_1$,\ldots,B$_{n}$\},
such that each reduced bipartite state
\begin{equation}
    \rho_{\text{AB}_i}:=\text{Tr}_{\text{B}_j\neq \text{B}_i}\rho_{\text{A}\textbf{B}^{(n)}}
\end{equation}
satisfies~$\rho_{AB_i}=\rho_\text{AB}$ for all $i=1,2,...,n$~\cite{MAG06}.
If $\rho_\text{AB}$ is $n$-shareable
for all $n$, then $\rho_\text{AB}$ is said to be $\infty$-shareable.
However, the only $\infty$-shareable states are separable
states~\cite{FLV88, RW89} so bipartite entanglement
cannot be shared among arbitrarily many parties whereas
some mixed-state entanglement can be shared simultaneously
for a finite number of parties~\cite{Ter04}.

There is no classical counterpart to this restricted sharability of correlations as all classical probability
distributions can be shared among all parties, and correlation between parties~A and~B (whether they are perfectly correlated
or not) does not restrict~A's correlation with~C. This
restricted sharability in quantum physics makes MoE fundamentally
different form classical physics. Throughout this article, we refer
to this restricted sharability of entanglement (both pure and
mixed) in multipartite systems as the monogamy of entanglement.

\section{Bipartite entanglement measures}
\label{sec:bipartite}

Whereas MoE is a property of \emph{multipartite} quantum
entanglement, it is expressed in terms of \emph{bipartite} measures
of entanglement~(\ref{monog}). In this section we consider choices
of bipartite measures of entanglement and their suitability for MoE.
In particular we consider bipartite entanglement measures for both
pure and mixed states and discuss analytical evaluation of the
entanglement measures for certain cases of low-dimensional quantum
systems. We refer the reader to~\cite{PV98} for a comprehensive
review of bipartite entanglement measures.

\subsection{Entanglement of bipartite pure states}
\label{subsec:Epure}

In this section we discuss how to quantify the entanglement between two parties
A and B who share a pure state $\rho_\text{AB}=|\psi\rangle_\text{AB}\langle\psi|$.
As entanglement cannot increase by LOCC, it must remain unchanged under \emph{reversible} LOCC.
An example of reversible LOCC is a local rotation or local unitary operation.
To illustrate, suppose A rotates her share of the system by some angle around a particular axis.
Such a rotation is reversible and is described mathematically by a unitary operator $U_\text{A}$.

The state of her system after rotation is $U_\text{A}\otimes
\openone_\text{B}|\psi\rangle_\text{AB}$. As this local rotation is
reversible, the entanglement of the state before the rotation must
be equal to the entanglement of the state after the rotation. More
generally, given two unitary matrices $U_\text{A}$ and $U_\text{B}$,
the entanglement of $|\psi\rangle_\text{AB}$ is the same as the
entanglement of $U_\text{A}\otimes
U_\text{B}|\psi\rangle_\text{AB}$. Hence, entanglement is quantified
by functions that are invariant under local unitary operations.

For pure states, this invariance under local unitary operations implies that
the degree of entanglement for $\rho_\text{AB}=|\psi\rangle\langle\psi|$
can be understood in terms of the entropy of its reduced state either over~A or over~B;
it does not matter which reduction as both reductions lead to the same result for entanglement.
The reduced state of a bipartite state~$\rho_\text{AB}$ is its partial trace over one of the two subsystems~A or~B
denoted
\begin{equation}
    \rho_\text{A}=\text{Tr}_\text{B}\rho_\text{AB}
\end{equation}
(for a partial trace over~B).
The entropy of the reduced state is denoted~$S(\rho_\text{A})$.

Many classical entropy functions are available~\cite{vNe27,vNe32,
Ren60,Tsa88,BS03}, which leads to a multitude of general quantum
entropies~$S(\rho)=\text{tr}[\rho f(\rho)]$ with function~$f$ quite
general. The choice of entropy function for entanglement evaluation
has an impact on the resultant evaluation of entanglement of the
state and therefore must be chosen judiciously.

An easy-to-calculate yet meaningful entropy is a linear function of the state `purity'~$\text{Tr}\rho^2$, namely
\begin{equation}
\label{eq:linen2}
    S_\text{lin}\left(\rho\right)=2\left(1-\text{tr}\rho^2\right).
\end{equation}
This entropy function corresponds to choosing $f(\rho)\equiv\rho$,
which is useful as the lowest-order approximation to general~$S_f$.
The purity has a maximum value of~$1$ for a pure (rank-one) reduced
state~$\rho_\text{A}$ and a minimum value of $1/d$ for the
completely mixed state~$\rho_\text{A}=\frac{1}{d}\openone_\text{A}$
for~$\openone$ the identity matrix acting on a $d$-dimensional
Hilbert space~$\mathscr{H}_d$.

As an example, the purity of the reduced one-party state obtained
from the singlet state~$\ket{\Psi^-}_\text{AB}$ (\ref{eq:singlet})
is
\begin{align}
\label{eq:singletreduced}
    \rho_\text{A}
        =&\text{tr}_\text{B} \ket{\Psi^-}_\text{AB}\bra{\Psi^-}
            \nonumber   \\
        =&\frac{1}{2}\left( \ket{0}_\text{A} \bra{0}+\ket{1}_\text{A}\bra{1}\right)
            \nonumber   \\
        =&\frac{1}{2}\openone_\text{A},
\end{align}
for~$\openone_A$ the identity operator acting on subsystem~A. The
maximally entangled state~$\ket{\Psi^-}_\text{AB}$ reduces to the
maximally-mixed state~$\frac{1}{2}I$ over either subsytem. In a
sense, all the information that describes~$\ket{\Psi^-}_\text{AB}$
is contained in entanglement or quantum correlations between
subsystems~A and~B.

Purity clarifies the absence of entanglement in the product state~$\ket{00}_\text{AB}$:
the reduced state
\begin{equation}
    \rho_\text{A}=\text{Tr}(\ket{00}_\text{AB}\bra{00})=\ket{0}_\text{A}\bra{0}
\end{equation}
has a purity of one, and the same relation holds for subsystem~B.
Therefore, entanglement or correlation cannot exist in this
bipartite state. The entropic
measure~$S_\text{lin}\left(\rho\right)$ (\ref{eq:linen2}) quantifies
the uncertainty associated with the eigenvalue spectrum of~$\rho$.
In particular, the entropy vanishes if~$\rho$ is a pure state, and,
for a two-dimensional system (i.e.\ \ qubit), this measure attains its
maximum value for the maximally-mixed state~$\openone/2$.

These examples of the singlet and product states illustrate the
quantitative relation between the entanglement of a bipartite pure
state and the purity of the states of subsystems.
The entanglement of a bipartite pure state~$\ket{\psi}_\text{AB}$ is maximal if the
purity of its subsystems represented by the reduced density operator
$\rho_\text{A}$ or $\rho_\text{B}$ is minimal. The pure state
$\ket{\psi}_\text{AB}$ has no entanglement if~$\rho_\text{A}$ and
$\rho_\text{B}$ are maximally pure, i.e.\ \ pure states. 
Furthermore, we also obtain an intuitive way to quantify the entanglement of a
bipartite pure state by quantifying the purity of its reduced
density matrices; less purity of reduced density matrices implies
more entanglement between subsystems. The entanglement between
subsystems~A and~B for a pure state~$\ket{\psi}_\text{AB}$ is
inversely proportional to the purity of the subsystems.

Linear entropy can be used to quantify the entanglement of a bipartite pure state.
We use the \emph{tangle}
\begin{equation}
\label{eq:Ptangle}
    \tau\left(\ket{\psi}_\text{AB}\bra{\psi}\right):=S_\text{lin}(\rho_\text{A}),
\end{equation}
with the reduced density operator
$\rho_\text{A}=\text{Tr}_\text{B}\ket{\psi}_\text{AB}\bra{\psi}$~\cite{Woo98}.
This measure yields~$0$ for product states and~$1$ for a maximally-entangled two-qubit state.
Furthermore, it can be shown that this measure cannot increase on average under LOCC~\cite{Vid00},
which is expected for a proper measure of entanglement.
That is, entanglement or quantum correlations should not increase by local means.
In the following subsection, we generalize this pure-state measure of entanglement to the case of mixed states.
In general, mixed-state entanglement is hard to calculate, but we will see that the generalization of the tangle to mixed two qubit states can be computed analytically.

\subsection{Entanglement of bipartite mixed states}
\label{subsec:Emixed}

In Subsec.~\ref{subsec:Epure} we studied tangle for pure bipartite states.
In this subsection we extend this analysis in a natural way to the case of mixed bipartite states.
Any bipartite mixed state $\rho_\text{AB}$ is Hermitian  and therefore can be written as the sum
\begin{equation}
\label{eq:rhosum}
    \rho_\text{AB}=\sum_i p_i|\psi_i\rangle_\text{AB}\langle\psi_i|\;,
\end{equation}
Intuitively the mixed state~$\rho_\text{AB}$ is a mixture of pure states
\begin{equation}
    \{|\psi_i\rangle_\text{AB},p_i\}
\end{equation}
with $|\psi_i\rangle_\text{AB}$ a pure bipartite state and~$p_i$ its probability.

This decomposition is not unique as another mixture of pure states can lead to the exact same state $\rho_\text{AB}$.
Therefore, all decompositions $\{|\psi'_j\rangle_\text{AB},p'_j\}$ for which
\begin{equation}
     \rho_\text{AB}=\sum_j p'_j|\psi'_j\rangle_\text{AB}\langle\psi'_j|
\end{equation}
are equally valid. We note that even the cardinality of different decompositions is not the same in general.

For each pure state~$|\psi_i\rangle$ we know the tangle $\tau(|\psi_i\rangle\langle\psi_i|)$ (\ref{eq:Ptangle}).
If we regard the mixed state~$\rho$ as an ensemble of pure states, each with known tangle,
we can characterize the tangle
of the mixed state as some function of the collection of tangles of elements in the ensemble.
As an example, consider the average tangle in the ensemble $\{|\psi_i\rangle_\text{AB},p_i\}$:
\begin{equation}
    \sum_i p_i\tau\left(\ket{\psi_i}_\text{AB}\bra{\psi_i}\right).
\label{eq:avetangle}
\end{equation}
As the decomposition of~$\rho$ into ensemble of pure states is not unique,
the expectation value of tangle depends on which ensemble is selected.
Our concern is with forming entangled states in the least expensive way possible.
The relevant expectation value of tangles is the infimum of the average entanglement over all decompositions.

Of course the question arises as to whether, for general~$\rho_\text{AB}$, there even exists a decomposition of~$\rho$ such that its average entanglement equals the infimum.
The answer to this question is an emphatic yes because the set of all ensembles for~$\rho$ is compact, and average tangle is a continuous function defined on the set of ensembles.
Although the decomposition with the minimum average entanglement is known to exist,
minimization is a NP-hard problem and hence intractable in general.
Fortunately, for two-qubit mixed states, an analytical formula is known~\cite{Woo98}.

We illustrate now how different the expectation value of the tangle can be for different ensembles by considering two extreme examples.
One example has a tangle of zero and the other a tangle of one for the same state.
Consider the two-qubit maximally-mixed state
\begin{align}
    \rho_\text{AB}
        =&\frac{1}{4}\openone_{AB}
            \nonumber   \\
        =&\frac{1}{4}
        \Big(\ket{00}_\text{AB}\bra{00}+\ket{01}_\text{AB}\bra{01}
            \nonumber   \\  &
        +\ket{10}_\text{AB}\bra{10}+\ket{11}_\text{AB}\bra{11}\Big),
\label{eq:2mmixed}
\end{align}
with
\begin{equation}
    \{\ket{00}_\text{AB}, \ket{01}_\text{AB}, \ket{10}_\text{AB}, \ket{11}_\text{AB}\}
\end{equation}
the standard orthonormal `computational' basis for the two-qubit system~AB.

As each pure state in the decomposition is a product state, the
average tangle of the decomposition~(\ref{eq:2mmixed}) is trivially zero.
Expressed in the two-qubit systems Bell basis~$\{\ket{\Psi^\pm},\ket{\Phi^\pm}:=\ket{00}\pm\ket{11}\}$,
the maximally-mixed state~(\ref{eq:2mmixed}) can be written as
\begin{align}
    \rho_\text{AB}
        =&\frac{1}{4}\openone_{AB}
                    \nonumber   \\
        =&\frac{1}{4}
            \Big(\ket{\Phi^+}_\text{AB}\bra{\Phi^+}+\ket{\Phi^-}_\text{AB}\bra{\Phi^-}
                    \nonumber   \\
            &+\ket{\Psi^+}_\text{AB}\bra{\Psi^+}+
    \ket{\Psi^-}_\text{AB}\bra{\Psi^-}\Big).
\label{eq:2mmixedb}
\end{align}
However, each Bell state in this decomposition is a two-qubit
maximally entangled state so the average tangle of this
decomposition is maximal, which is markedly different from the case
of Eq.~(\ref{eq:2mmixed}). Thus the average tangle in
Eq.~(\ref{eq:avetangle}) is not always unique for a bipartite mixed state,
and average tangle strongly depends on which pure-state ensemble we choose to
represent $\rho_\text{AB}$.

Now consider the `cheapest' way to prepare the bipartite mixed state~(\ref{eq:rhosum}) by preparing
an ensemble of pure states~$\{p_i, \ket{\psi_i}_\text{AB}\}$ with~$p_i$ the
probability of preparing~$\ket{\psi_i}_\text{AB}$.
The least amount of resources required to prepare~$\rho_\text{AB}$
is the average tangle minimized over all distinct decompositions~$\{p_i, \ket{\psi_i}_\text{AB}\}$:
\begin{equation}
\label{eq:Mtangle}
    \tau\left(\rho_\text{AB}\right):=\min\sum_i p_i\tau\left(\ket{\psi_i}_\text{AB}\bra{\psi_i}\right).
\end{equation}
Eq.~(\ref{eq:Mtangle}) serves as the definition of the tangle for mixed state~$\rho_\text{AB}$,
and the procedure of minimizing over all pure-state decompositions to determine entanglement
is known as the `convex-roof extension'.

Many entanglement measures for bipartite
mixed states are based on this concept~\cite{Woo98, Ren60, Tsa88}.
For example, entanglement of formation of a bipartite mixed state~$\rho_\text{AB}$ is~\cite{BDSW96}
\begin{align}
    E_\text{f}(\rho_\text{AB})
        :=&\min\sum_i p_i S_\text{vN}\left(\rho_\text{A}^i\right),\,
        \rho_\text{A}^i
            \nonumber   \\
        :=&\text{tr}_\text{B} \ket{\psi_i}_\text{AB}\bra{\psi_i},\label{eof}
\end{align}
minimized over all pure-state ensembles,
where
\begin{equation}
\label{eq:vN}
    S_\text{vN}(\rho):=-\text{tr}\rho\log \rho
\end{equation}
is the von Neumann entropy of $\rho$~\cite{vNe32}.
As the convex-roof extension is typically hard to evaluate because minimizing over all all possible
pure-state decompositions is daunting,
linear entropy is preferred for qubits.
This preference arises because the qubit equations are tractable and even analytically solved~\cite{Woo98},
which is another reason why tangle, derived from linear entropy, is so prevalent.

We complete this section by presenting the formula for the tangle of two-qubit mixed states.
The formula is technical and is given in terms of the `spin-flipped' two-qubit mixed state
\begin{equation}
\label{eq:rhotilde}
    \tilde{\rho}_\text{AB}=(\sigma_y \otimes\sigma_y)
        \rho^*_{AB}(\sigma_y\otimes\sigma_y)
\end{equation}
with~$\rho_\text{AB}$ the actual two-qubit state,
$\rho_\text{AB}^*$ its complex conjugate with respect to the standard basis
and
\begin{equation}
    \sigma_y := \begin{pmatrix}0& -i\\i&0\end{pmatrix}
\label{eq:sigmay}
\end{equation}
a Pauli operator in the standard basis. Denoting by
$\{\lambda_i\}$ the set of eigenvalues, in decreasing order, of the Hermitian operator
\begin{equation}
    \sqrt{\sqrt{\rho_\text{AB}}\tilde{\rho}_\text{AB}\sqrt{\rho_\text{AB}}},
\end{equation}
the tangle of~$\rho_\text{AB}$ has then the analytical expression~\cite{Woo98}
\begin{equation}
    \tau(\rho_\text{AB})=\left(\lambda_1-\lambda_2-\lambda_3-\lambda_4\right)^2,
\label{t_formula}
\end{equation}
for
\begin{equation}
\label{eq:lambdas}
    \lambda_1\geq \lambda_2+\lambda_3+\lambda_4;
\end{equation}
otherwise, $\tau(\rho_\text{AB})=0$.

Other analytically evaluated entanglement measures including entanglement of formation
exist for two-qubit systems~\cite{Kim10}, but their analytic evaluation is actually based on that of the tangle.
In other words, the tangle is the only known bipartite entanglement measure whose evaluation
for two-qubit mixed states is analytical.
Wootters's detailed proof of the analytic formula is technical and not shown here~\cite{Woo98}.

\subsection{Summary}
In closing this section,
we have shown that various entanglement measures are possible for bipartite systems.
Furthermore, entanglement of a mixed system, expressed as a function of the average entanglement of ensemble of pure states obtained
via a pure-state decomposition of the mixed state, is not unique. One of the reasons for that is that the pure-state decomposition of a mixed state is not unique.
For qubit systems the tangle, derived from linear entropy, is especially appealing as an entanglement measure because of its amenability to analytical evaluation.
The mixed-state tangle is the average tangle for the pure-state decomposition that uniquely minimizes this function.

The tangle and the linear entropy play a key role in the next section, which explains the MoE for three-qubit systems.
Although linear entropy can be extended from qubits to higher dimensions, we see in subsequent sections that extensions of monogamy relations using linear entropy for higher dimensions is problematic.

\section{Monogamy of entanglement in three-qubit systems }
\label{Sec: Monoqubit}

In this section we consider a three-party system ABC with each party holding a qubit.
Earlier we saw that, if any two parties hold a pure maximally entangled state, the third party cannot share any entanglement with the first two.
Our argument was based on teleportation and no cloning principles.
Now we show that entanglement cannot be shared but this time using the concept of entanglement measures.

If the two parties sharing composite subsystem AB hold a pure entangled state $\ket{\psi}_\text{AB}$,
then the linear entropy of this state is
$S_\text{lin}\left(\ket{\psi}_\text{AB}\bra{\psi_\text{AB}}\right)=0$.
The tripartite state~$\rho_\text{ABC}$ must be a product state,
$\rho_\text{ABC}=\ket{\psi}_\text{AB}\bra{\psi_\text{AB}}\otimes\rho_\text{C}$ for AB to share a zero-entropy state.
Therefore, if two parties share a pure maximally entangled state, then there cannot be any entanglement shared between AB and anyone else,
hence the concept of MoE.
On the other hand, if AB share an entangled state that is not maximal,
A or B or both could share some entanglement with other parties, but the amount of allowed shared entanglement is bounded.

Limited sharability of entanglement between parties is quantified by monogamy-of-entanglement mathematical relations.
In this section, we discuss restricted sharability of pure and mixed bipartite entanglement for just three qubits shared by three separate parties.

\subsection{From W versus GHZ interpolation to monogamy inequality of multi-qubit entanglement}
\label{Sec: W vs GHZ}

In Sec.~\ref{sec:entanglement} we examined the pure three-qubit W state~(\ref{eq:W}) and GHZ state~(\ref{eq:GHZ}).
All three-qubit states are equivalent to one or the other of these states under SLOCC.
We start this section by showing that both the W and GHZ states are maximally entangled but in distinct ways:
the W state maximizes the expected two-qubit bipartite entanglement after tracing out the third qubit
whereas the GHZ state maximizes the expected bipartite entanglement between one qubit
and the other two qubits. We will see that a comparison between these two distinct ways to quantify the averaged tripartite entanglement
leads to a monogamy inequality that can be generalized to any number of qubits/parties.

By using the tangle to quantify the bipartite entanglement between two parties,
we can quantify the expected entanglement rigorously.
Consider a general three-qubit pure state
\begin{equation}
\label{eq:threequbitpurestate}
    \ket{\phi}_\text{ABC}\in \mathbb{C}^2\otimes \mathbb{C}^2\otimes\mathbb{C}^2
\end{equation}
and denote its bipartite entanglement with respect to the three possible bi-partitions by
\begin{equation}
    \tau_\text{A$|$BC},\;
    \tau_\text{B$|$AC},\;
    \tau_\text{C$|$AB}.
\end{equation}
For example $\tau_\text{A$|$BC}$ is brief notation for $\tau_\text{A$|$BC}\left(\ket{\phi}_\text{ABC}\bra{\phi}\right)$, which measures the tangle between qubit~A and the 2-qubit system~BC.
With this notation we define the expectation value of the tangle, denoted here by
\begin{equation}
\label{eq:3avgT}
    \tau_1\left(\ket{\phi}_\text{ABC}\bra{\phi}\right):=\frac{\tau_\text{A$|$BC}
    +\tau_\text{B$|$AC}+
    \tau_\text{C$|$AB}}{3}
\end{equation}
with each bipartition allotted equal importance (i.e.\ , weight).
The average tangle $\tau_1$ is always non negative,
and,
for three-qubit systems,
the tangle cannot exceed unity.
This maximal value is achieved by the GHZ state, i.e.\ ,
\begin{equation}
\label{eq:tau1GHZ}
    \tau_1\left( \ket{\text{GHZ}}_\text{ABC}\bra{\text{GHZ}}\right)=1.
\end{equation}

On the other hand,
we can define another averaged value for the tangle of three qubits, denoted here by $\tau_2$, which is based on the two-qubit tangle
inherent in a three-qubit state~$\ket{\phi}_\text{ABC}$; i.e.\
\begin{equation}
    \tau_2\left(\ket{\phi}_\text{ABC}\bra{\phi}\right):=\frac{\tau_\text{A$|$B}
    +\tau_\text{B$|$C}+\tau_\text{A$|$C}}{3},
\label{eq:T_2}
\end{equation}
where~$\tau_\text{A$|$B}$ (and similarly $\tau_\text{B$|$C}$ and $\tau_\text{A$|$C}$) is a short notation for $\tau\left(\rho_\text{AB}\right)$ where $\rho_\text{AB}$ (and similarly $\rho_\text{BC}$ and~$\rho_\text{AC}$) is the
two-qubit reduced density matrix of~$\ket{\phi}_\text{ABC}$ obtained after tracing out system~C.
For the GHZ state
\begin{equation}
    \tau_2\left(\ket{\text{GHZ}}_\text{ABC}\bra{\text{GHZ}}\right)=0,
\label{eq:T_2GHZ3}
\end{equation}
because every two-qubit reduced density matrix of
$\ket{\text{GHZ}}_\text{ABC}$ is separable.
Thus, the averaged two-qubit entanglement in~$\ket{\text{GHZ}}_\text{ABC}$ is nil despite
its expected entanglement with respect to every bipartition being maximal.

Classically any correlation between subsystem~A
and composite subsystem BC consists of the correlation of
$A$ with each of~B and~C. If neither~B nor~C is correlated with
$A$, then there is no correlation between~A and~BC.
However, the GHZ state~(\ref{eq:T_2GHZ3}) provides a counter-example to this rule
in the quantum case.
Quantum entanglement between system~A and composite system BC
does not guarantee the entanglement between~A and~B or between~A and~C.

In fact the GHZ state is an extreme quantum state whose expected bipartite entanglement between one
qubit and two qubits is maximal (unity) whereas its expected two-qubit
entanglement inherent is minimal (nil):
\begin{align}
    1=&\tau_1\left(\ket{\text{GHZ}}_\text{ABC}\bra{\text{GHZ}}\right)
                \nonumber   \\
    >&\tau_2\left(\ket{\text{GHZ}}_\text{ABC}\bra{\text{GHZ}}\right)
                \nonumber   \\
    =&0.
\label{T2T3GHZ}
\end{align}
Furthermore, the inequality between $\tau_1$ and~$\tau_2$~(\ref{T2T3GHZ}) is always valid for any pure three-qubit state $\ket{\phi}_\text{ABC}$.
For a general three-qubit state $\ket{\phi}_\text{ABC}$, each two-qubit reduced density matrix is
obtained by tracing out a one-qubit subsystem. Due to the monotonicity
of entanglement, discarding subsystems does not increase entanglement so
\begin{align}
    \tau_\text{A$|$BC}\geq \tau_\text{A$|$B},~
    \tau_\text{B$|$CA}\geq \tau_\text{B$|$C},~
    \tau_\text{C$|$AB}\geq \tau_\text{C$|$A}
\label{eq:reducrel}
\end{align}
for the two-qubit reduced density matrices~$\rho_\text{AB}$, $\rho_\text{BC}$
and~$\rho_\text{AC}$.
Relations~(\ref{eq:reducrel}) imply that
\begin{equation}
    \tau_1\left(\ket{\phi}_\text{ABC}\bra{\phi}\right)\geq\tau_2\left(\ket{\phi}_\text{ABC}\bra{\phi}\right),
\label{eq:t2t3gen1}
\end{equation}
for any three-qubit state~$\ket{\phi}_\text{ABC}$.
We also observe that the GHZ state assumes the largest difference between the two terms
in Inequality~(\ref{eq:t2t3gen1}).

Inequality~(\ref{eq:t2t3gen1}) provides an upper bound for the
simultaneously shared two-qubit entanglement
$\tau_2\left(\ket{\phi}_\text{ABC}\bra{\phi}\right)$ in terms
of~$\tau_1\left(\ket{\phi}_\text{ABC}\bra{\phi}\right)$ for a three-qubit state~$\ket{\phi}_\text{ABC}$.
However, this inequality is never saturated by
any three-qubit entangled state:
its equality holds only if
$\ket{\phi}_\text{ABC}$ is a three-way product state.

Although $\tau_1\left(\ket{\phi}_\text{ABC}\bra{\phi}\right)$ can
assume an arbitrary value between $0$ and~$1$,
$\tau_2\left(\ket{\phi}_\text{ABC}\bra{\phi}\right)$ cannot equals unity.
If
\begin{equation}
    \tau_2\left(\ket{\psi}_\text{ABC}\bra{\psi}\right)=1,
\end{equation}
this equality would imply that all two-qubit reduced density matrices
$\{\rho_\text{AB},\rho_\text{BC},\rho_\text{AC}\}$ are maximally entangled.
Such simultaneous
maximal entanglement would enable~A to
teleport an arbitrary unknown state to~B and to~C simultaneously
and thereby contradict the no-cloning principle of quantum mechanics~\cite{WZ82}.

The averaged tangle $\tau_2\left(\ket{\phi}_\text{ABC}\bra{\phi}\right)$ in~(\ref{eq:t2t3gen1})
is maximized by the W state:
\begin{equation}
    \tau_2\left(\ket{\phi}_\text{ABC}\bra{\phi}\right)\leq\tau_2\left(\ket{\text{W}}_\text{ABC}\bra{\text{W}}\right)=\frac{4}{9}\;.
\label{eq:t2W}
\end{equation}
This upper bound follows from the closed formula in Eq.~(\ref{t_formula}) for the tangle between two qubits in a mixed state and reveals the mutually-exclusive nature of mixed-state monogamy
in tripartite quantum systems.
In three-qubit systems $ABC$, if
the amount of entanglement between two parties increases,
then the amount of entanglement between other parties must
decrease so that the sum of all possible two-qubit
entanglement does not exceed the upper bound~(\ref{eq:t2W}).

This mutually exclusive relation for two-qubit
entanglement within three-qubit systems is most prominent for states in the $W$~class under SLOCC.
States in the W class have the form
\begin{equation}
    \ket{\phi}_\text{ABC}=a\ket{100}_\text{ABC}+b\ket{010}_\text{ABC}+c\ket{001}_\text{ABC}
\label{3Wclass}
\end{equation}
with $a,~b,~c \in {\mathbb C}$ satisfying $|a|^2 +|b|^2 +|c|^2 =1$.
The `W state' is a special case in which $a=b=c=1/\sqrt{3}$.
For these $W$-type states, the closed formula in Eq.~(\ref{t_formula}) implies that
\begin{equation}
    \tau_\text{A$|$BC}=\tau_\text{A$|$B}+\tau_\text{A$|$C}.
\label{Wclassequal}
\end{equation}
That is, W-type states saturate the monogamy relation~\eqref{monog}.
Additionally, as W-type states are symmetric under permutation of the qubits,
Eq.~(\ref{Wclassequal}) also holds under any permutation of~A,~B, and~C.   Therefore, the above equality also implies that
for a three-qubit W-class state~$\ket{\phi}_\text{ABC}$ under SLOCC, we have
\begin{equation}
    \tau_1\left(\ket{\phi}_\text{ABC}\bra{\phi}\right)=2\tau_2\left(\ket{\phi}_\text{ABC}\bra{\phi}\right).
\label{eq:t2t3Wgen}
\end{equation}

\begin{figure}
\includegraphics[width=8cm]{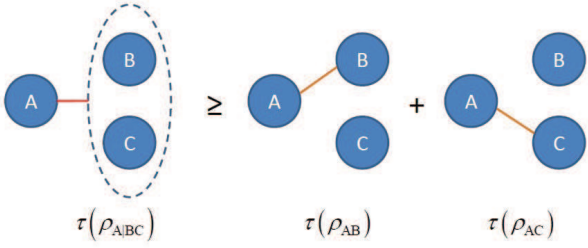}
\caption{
    Characterization of bipartite entanglement shared in three-qubit systems:
    the Coffman-Kundu-Wootters inequality.
    }
\label{fig:2qubitint2qubitsystem}
\end{figure}

The first characterization of MoE in three-qubit systems was
introduced by extending~(\ref{Wclassequal}) into an inequality for a
general three-qubit pure state~$\ket{\psi}_\text{ABC}$.
Specifically Coffman-Kundu-Wootters showed that
\begin{equation}
\tau_\text{A$|$BC}\geq\tau_\text{A$|$B}+\tau_\text{A$|$C}
\label{TCKWpure}
\end{equation}
for any three-qubit state~$\ket{\psi}_\text{ABC}$~\cite{CKW00},
which is depicted in Fig.~\ref{fig:2qubitint2qubitsystem}.
This Coffman-Kundu-Wootters inequality implies that Inequality~(\ref{eq:t2t3gen1}),
which never saturates for three-qubit entangled
states, tightens to
\begin{align}
    \tau_1\left(\ket{\psi}_\text{ABC}\bra{\psi}\right)\geq2\tau_2\left(\ket{\psi}_\text{ABC}\bra{\psi}\right)
\label{T2T3mono}
\end{align}
for any three-qubit state~$\ket{\psi}_\text{ABC}$ and is saturated for W-type states.

Inequality~(\ref{TCKWpure}) provides a non-trivial upper bound for simultaneously shared two-qubit
entanglement in three-qubit systems.
Although subsystem~A can be simultaneously entangled with both~B and~C, the sum of each
entanglement cannot exceed the entanglement between~A and~BC.
Moreover, if~$\tau_\text{A$|$BC} =
\tau_\text{A$|$B}=1$, meaning that~A and~B share
maximal entanglement, Inequality~(\ref{TCKWpure}) implies~$\tau_\text{A$|$C}=0$.
Thus there cannot be any entanglement between~A and~C,
and Inequality~(\ref{TCKWpure}) reduces to the singlet monogamy relation discussed
in Sec.~\ref{sec:entanglement}.
Note that the Coffman-Kundu-Wootters inequality is
tight in a non-trivial sense because it is saturated by all
three-qubit W states~(\ref{3Wclass}).
This mutually exclusive relation of two-qubit entanglement
in the Coffman-Kundu-Wootters inequality is illustrated in Fig.~\ref{fig:2qubitint2qubitsystem} .

We end this section generalizing the Coffman-Kundu-Wootters inequality to three-qubit mixed states and to more than three qubits.
Consider a three-qubit mixed state
\begin{equation}
    \rho_\text{ABC}=\sum_i p_i \ket{\psi_i}_\text{ABC}\bra{\psi_i}.
\label{3mix}
\end{equation}
Let us assume that $\{p_i,\;\ket{\psi_i}_\text{ABC}\}$ is the optimal decomposition minimizing the average tangle of
$\rho_\text{ABC}$ with respect to the bipartite cut between~A and~BC.
In other words
\begin{equation}
\label{eq:tau_A|BC}
    \tau_\text{A$|$BC}\left(\rho_\text{ABC}\right)
        =\sum_i p_i \tau_\text{A$|$BC}\left(\ket{\psi_i}_\text{ABC}\bra{\psi_i}\right).
\end{equation}
As each~$\ket{\psi_i}_\text{ABC}$ in the decomposition is a
three-qubit pure state, it satisfies monogamy inequality~(\ref{TCKWpure}).
Thus, for each~$i$,
\begin{equation}
    \tau_\text{A$|$BC}\left(\ket{\psi_i}_\text{ABC}\bra{\psi_i}\right)
        \geq \tau\left(\rho^i_{AB} \right)+\tau\left(\rho^i_{AC} \right),
\label{Tmonoi}
\end{equation}
with~$\rho^i_{AB}$ and~$\rho^i_{AC}$ the reduced density
matrices of~$\ket{\psi_i}_\text{ABC}$ onto subsystems~AB and~AC, respectively.
Inequality~(\ref{Tmonoi}) and Eq.~(\ref{eq:tau_A|BC}) together yield
\begin{align}
    \tau_\text{A$|$BC}\left(\rho_\text{ABC}\right)\geq \tau\left(\rho_\text{AB}\right)+\tau\left(\rho_\text{AC}\right)
\label{3Tmonomixed}
\end{align}
for a three-qubit mixed state~$\rho_\text{ABC}$ and its reduced density
matrices~$\rho_\text{AB}$ and~$\rho_\text{AC}$. Thus the Coffman-Kundu-Wootters inequality in
(\ref{TCKWpure}) is also true for three-qubit mixed states.

Inequality~(\ref{3Tmonomixed}) has been generalized for an arbitrary number of qubit-systems~\cite{OV06}.
This generalization is
\begin{equation}
\label{eq:nTmonomixed}
    \tau\left(\rho_{\text{A}_1|\text{A}_2\cdots\text{A}_n}\right) \geq
        \tau\left(\rho_{\text{A}_1\text{A}_2} \right)+\cdots +
        \tau\left(\rho_{\text{A}_1\text{A}_n}\right),
\end{equation}
for any $n$-party state~$\rho_{\text{A}_1\text{A}_2\cdots\text{A}_n}$ with
$\rho_{\text{A}_1\text{A}_i}$ the reduced density matrix
acting on subsystems $\text{A}_1\text{A}_i$
for $i=2,\ldots, n$.

\subsection{Polygamy: dual monogamy inequality in multi-party quantum systems}
\label{Sec: Dual}

In this section we discuss a concept that is dual to monogamy of entanglement.
In particular we shall see that,
whereas \emph{sharing} entanglement is monogamous, \emph{distributing} entanglement is polygamous.
Consider a bipartite mixed state~$\rho_\text{AB}$.
Any mixed state can be considered to be a reduced pure state acting over a larger
Hilbert space, and constructing such a pure state from the mixed state is known as a purification.
Here we denote the state arising from a purification of ~$\rho_\text{AB}$ as
$\ket{\psi}_\text{ABC}$ such that $\rho_\text{AB}=\text{tr}_\text{C}(\ket{\psi}_\text{ABC}\bra{\psi})$.
Party~C can help to increase the entanglement of
$\rho_\text{AB}$ by performing measurements on C's own subsystem and then
communicating the measurement results to~A and~B.

As an extreme example, consider the tripartite state
\begin{equation}
    \ket{\psi}_\text{ABC}\equiv\ket{\psi^+}_\text{AB}|0\rangle_\text{C}+\ket{\psi^-}_\text{AB}|1\rangle_\text{C}.
\end{equation}
A simple calculation shows that the state $\rho_\text{AB}=\text{tr}_\text{C}(\ket{\psi}_\text{ABC}\bra{\psi})$
is separable, so that parties~A and~B do not share entanglement. However, suppose party~C performs a measurement
in the $|0\rangle$ and $|1\rangle$ computational basis.
After such a measurement parties~A and~B will share the maximally entangled state $\ket{\psi^+}_\text{AB}$ or $\ket{\psi^-}_\text{AB}$ depending on the measurement outcome.
With the assistance of~C, parties~A and~B can end up with a maximally entangled state.
This ability of a party who holds a purification to assist, leads to the concept
of \emph{entanglement of assistance},
which is defined as the
maximum possible average entanglement that can be created between parties~A and~B
with the assistance of~C~\cite{LVvE03}.

A one-to-one correspondence between rank-one measurements of~C and
the pure-state ensembles of~$\rho_\text{AB}$ exists~\cite{LVvE03}.
This correspondence implies that the maximum average entanglement
that can be created between the parties~A and~B is given by
\begin{equation}
\label{eq:ToA}
    \tau^\text{a}\left(\rho_\text{AB}\right):=\max\sum_i p_i \tau(\ket{\psi_i}_\text{AB}\bra{\psi_i}),
\end{equation}
which is called the  \emph{tangle of assistance}, where the maximum is taken over all pure-state decompositions representing
$\rho_\text{AB}=\sum_ip_i\ket{\psi_i}_\text{AB}\bra{\psi_i}$.

The tangle of assistance is dual to the tangle~(\ref{eq:Mtangle}) with this duality mathematically clear because
the tangle of assistance is a maximum whereas the tangle is a minimum over all pure-state ensembles realizing $\rho_\text{AB}$. Furthermore, $\tau\left(\rho_\text{AB}\right)$
represents the minimum entanglement needed to create $\rho_\text{AB}$
(formation) whereas $\tau^\text{a}\left(\rho_\text{AB}\right)$ is the maximum
amount of averaged entanglement attainable between~A and~B with the assistance of~C.
Thus, the mathematical duality between the tangle and the tangle of assistance for bipartite mixed states has physical meaning.

Whereas monogamy inequalities~(\ref{eq:nTmonomixed}) provide an
\emph{upper} bound for the restricted sharability of multi-party
entanglement, this same bound serves as a \emph{lower} bound for the
distribution of entanglement of assistance in multi-party quantum systems.
In multi-qubit systems, entanglement of assistance is described mathematically  as a
dual inequality to~(\ref{eq:nTmonomixed}), namely~\cite{GBS07}
\begin{equation}
\label{eq:npoly}
    \tau\left(\rho_{\text{A}_1|\text{A}_2\cdots\text{A}_n}\right) \leq
    \tau^\text{a}\left(\rho_{\text{A}_1\text{A}_2} \right)+\cdots +
    \tau^\text{a}\left(\rho_{\text{A}_1\text{A}_n}\right).
\end{equation}
Inequality~(\ref{eq:npoly}) is also called the polygamy of entanglement for multi-qubit systems.

Recently a general polygamy-of-entanglement inequality was proposed for three-party quantum
systems with an arbitrary number of Hilbert-space dimensions using the von Neumann
entropy~\cite{BGK09}.
In particular, for any tripartite pure state
$|\psi\rangle_\text{ABC}$ in any number of dimensions,
\begin{equation}
\label{eq:PoEinequality}
    E_{\text{A}|\text{BC}}
        \equiv S_\text{vN}(\rho_\text{A})
        \leq E^\text{a}(\rho_\text{AB})
            +E^\text{a}(\rho_\text{AC})
\end{equation}
holds for~$E^\text{a}$ the entanglement of assistance that is dual to the entanglement of
formation
Thus, for a tripartite pure state of
arbitrary dimension, there exists a polygamy-of-entanglement relation in terms of entropy
of entanglement (i.e.\ \ $E_{\text{A}|\text{BC}}$) and entanglement of assistance.

Inequality~(\ref{eq:PoEinequality}) is
the first known result for the polygamous (or dually monogamous)
property of distribution of entanglement in multipartite higher
dimensional quantum systems other than qubits. In addition, a tight
upper bound on the polygamy-of-entanglement inequality has been proposed for an
arbitrary-dimensional multi-party quantum systems~\cite{Kim09}.
Recently, a general polygamy inequality of entanglement in a
multi-party quantum system of arbitrary dimensions has been
established in terms of entanglement of assistance~\cite{Kim12}.

\section{Monogamy of entanglement in higher-dimensional quantum systems}
\label{Sec: Higher-dimensional monogamy}

\subsection{Entanglement for higher-dimensional systems}
\label{subsec:entanglementhigher}

Generalization of monogamy of entanglement from the multi-qubit case to the multi-qudit case
is complicated by the wealth of possible entanglement measures and the specifics
drawbacks that each such measure entails.
For pure bipartite states, these measures are constructed from R\'enyi-$\alpha$ and Tsallis-$q$
entropies (for appropriate values of~$\alpha$ and~$q$)~\cite{Ren60, Tsa88}.

For any quantum state $\rho$, the quantum R\'enyi entropy of order $\alpha$ (or R\'enyi-$\alpha$ entropy) is
\begin{equation}
	S_{\alpha}(\rho):=\frac{1}{1-\alpha}\log \text{tr} \rho^{\alpha},
\label{r-entropy}
\end{equation}
for any $\alpha >0$ and $\alpha \neq 1$. Similarly, the Tsallis-$q$ entropy of a quantum state $\rho$ is defined as
\begin{equation}
	T_{q}(\rho):=\frac{1}{q-1}\left(1- \text{tr} \rho^{q}\right),
\label{q-entropy}
\end{equation}
for any $q >0$ and $q \neq 1$.
In the limitimg case where $\alpha \rightarrow 1$ and $q \rightarrow 1$, R\'enyi-$\alpha$ and Tsallis-$q$ entropies
converge to the von Neumann entropy respectively, that is, R\'enyi-$\alpha$ and Tsallis-$q$ entropies are one-parameter generalizations of von Neumann entropy.
As for the tangle whose derivation is based on the linear entropy,
R\'enyi-$\alpha$ and Tsallis-$q$ entropies also lead us to various classes of entanglement measure for bipartite pure states,
as well as the extended to mixed states via the convex-roof extension.

Analytic evaluations of these entropy-based entanglement measures for a
two-qubit state~$\rho_\text{AB}$ are based on the existence of a pure-state decomposition of~$\rho_\text{AB}$ that realizes simultaneously the minimum average entanglement
for {\it all} these entanglement measures.
If the decomposition $\rho_\text{AB}=\sum_ip_i \ket{\psi_i}_\text{AB}\bra{\psi_i}$
realizes the minimum average tangle such that
\begin{equation}
    \tau\left(\rho_\text{AB}\right)=\sum_ip_i \tau\left(\ket{\psi_i}_\text{AB}\bra{\psi_i}\right),
\end{equation}
then it also realizes the
minimum average entanglement for other entropy-based entanglement
measures. Moreover, these minimum values are monotonically related
to each other by a continuous function~\cite{Kim10, KS10}.
In other words the analytic evaluation of these measures are directly derived from
that of the tangle and thus are all equivalent to each other because
they can be obtained from the two-qubit tangle via a continuous function.
In this sense, we observe that the tangle is the only known bipartite
entanglement measure so far whose analytic evaluation is tractable
in two-qubit systems.

\subsection{Generalization of the Coffman-Kundu-Wootters inequality}
\label{subsec:generalization}

Now we consider a possible generalization of the Coffman-Kundu-Wootters
inequality into higher-dimensional quantum systems.
As the tangle~(\ref{eq:Mtangle}) is well-defined for bipartite quantum states of arbitrary dimension,
we may expect the validity of the Coffman-Kundu-Wootters inequality itself in higher dimensions.
However, there exist counter-examples that violate the Coffman-Kundu-Wootters inequality for tripartite quantum systems
where any single subsystem has more than two dimensions~\cite{Ou07,KS08}.

We must conclude that the Coffman-Kundu-Wootters inequality only holds
in terms of two-qubit tangles for multi-qubit systems.
Even a tiny extension for any subsystem leads to a violation. For the generalization of the Coffman-Kundu-Wootters inequality
to higher dimensional quantum systems, we thus need to consider
other possible entanglement measures.

\subsection{Squashed entanglement: an additive and monogamous entanglement measure}
\label{subsec:squashed}

In this subsection we elaborate on a measure of entanglement that is monogamous,
i.e.,  satisfies Eq.~\eqref{monog}
for all dimensions. The existence of such a measure is important as it shows that entanglement is truly monogamous in all dimensions. In order to understand the significance of the new measure, we first discuss the intimate relationship between monogamy of entanglement in higher dimensions and another desired property of
entanglement: additivity under tensor products.

An additive entanglement measure~$E$ satisfies the property
\begin{equation}
\label{eq:add}
    E\left(\rho_{\text{A}_1\text{B}_1}\otimes \sigma_{\text{A}_2\text{B}_2} \right)=
    E\left(\rho_{\text{A}_1\text{B}_1}\right)+E\left(\sigma_{\text{A}_2\text{B}_2} \right)
\end{equation}
for any bipartite state~$\rho_{\text{A}_1\text{B}_1}$ and~$\sigma_{\text{A}_2\text{B}_2}$.
Additivity is a desirable entanglement-measure property in connection
with an operational interpretation as an interconversion rate for a quantum information processing task.
In such cases, the rate of the quantum information processing task is given by the regularized version of some entanglement measure,
namely,
\begin{equation}
    R=\lim_{n\to\infty}\frac{E\left(\rho_\text{AB}^{\otimes n}\right)}{n}\;,
\end{equation}
with~$E$ an entanglement measure.

If~$E$ is additive, then the simple relation $R=E$ holds.
Otherwise, $R$ is difficult to calculate.
For example, the entanglement cost, which quantifies the rate at which many Einstein-Podolsky-Rosen pairs can be converted to many copies of a bipartite mixed state $\rho_\text{AB}$ by LOCC,
is given in terms of the regularized version of the entanglement of formation of $\rho_\text{AB}$.
Recently, Hastings discovered that the entanglement of formation is not additive~\cite{Has09}, which makes calcuatling the entanglement cost difficult.

In higher dimensions, monogamy inequalities lead to additivity or strong superadditivity of entanglement measures~$E$.
A strongly-superadditive entanglement measure~$E$ satisfies the property
\begin{align}
    E\left(\rho_{\text{A}_1\text{A}_2|\text{B}_1\text{B}_2}\right)
    \geq& E\left(\rho_{\text{A}_1\text{B}_1}\right) + E\left(\rho_{\text{A}_2\text{B}_2}\right)
\label{eq:SS}
\end{align}
for a four-party state~$\rho_{\text{A}_1\text{A}_2\text{B}_1\text{B}_2}$.

To illustrate this requirement of strong superadditivity,
first consider the two subsystems $\text{A}_1\text{A}_2$ of the four-party state~$\rho_{\text{A}_1\text{A}_2\text{B}_1\text{B}_2}$ 
treated as a single party.
Monogamy implies that
\begin{align}
    E\left(\rho_{\text{A}_1\text{A}_2|\text{B}_1\text{B}_2}\right)\geq & E\left(\rho_{\text{A}_1\text{A}_2|\text{B}_1}\right)
    +E\left(\rho_{\text{A}_1\text{A}_2|\text{B}_2}\right)\nonumber\\
    \geq& E\left(\rho_{\text{A}_1\text{B}_1}\right) + E\left(\rho_{\text{A}_2\text{B}_2}\right),
\label{eq:SSmono1}
\end{align}
which is strongly superadditive for~$E$.

\begin{figure}
\includegraphics[width=8cm]{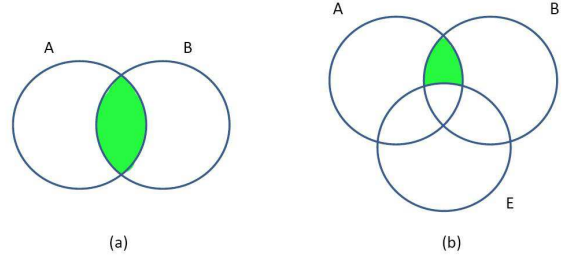}
\caption{
    Venn diagrams of (a)~the mutual information $I(A;B)$
    and (b)~the conditional mutual information $I(A;B|E)$.
    }
\label{fig:Venn}
\end{figure}

Only two entanglement measures are known to be additive:
squashed entanglement~\cite{CW04} and logarithmic negativity~\cite{ZPSL98,VW02}.
Squashed entanglement is especially appealing because it is also monogamous for quantum systems of arbitrary dimension.
For the state `extension'
\begin{equation}
\label{eq:ext}
    \text{ext}\rho_\text{AB}:=\left\{\rho_\text{ABE};
        \text{tr}_\text{E}\rho_\text{ABE}=\rho_\text{AB}\right\}
\end{equation}
being the set of all ABE mixed states in any dimension that are compatible with the joint AB state under a partial trace,
the squashed entanglement for such a bipartite state~$\rho_\text{AB}$ is
\begin{align}
\label{eq:squashed}
    E_\text{sq}\left(\rho_\text{AB}\right)
        :=\frac{\inf_{\rho_\text{ABE}\in\text{ext}\rho_\text{AB}} \left\{I(A;B|E)\right\}}{2}
\end{align}
with~$I(A;B|E)$ the quantum conditional mutual information of~$\rho_\text{ABE}$~\cite{CA97},
namely,
\begin{align}
    I(A;B|E):=&S_\text{vN}(\rho_\text{AE})+S_\text{vN}(\rho_\text{BE})\nonumber\\
           &-S_\text{vN}(\rho_\text{ABE})-S_\text{vN}(\rho_\text{E}).
\end{align}
 The relationship between mutual information and tripartite
conditional mutual information is depicted in Fig.~\ref{fig:Venn}.

For a bipartite pure state~$\ket{\psi}_\text{AB}$, any possible extension $\rho_\text{ABC}$
such that
\begin{equation}
    \text{tr}_\text{C} \rho_\text{ABC}=\ket{\psi}_\text{AB}\bra{\psi}
\end{equation}
must be a product state
\begin{equation}
    \ket{\psi}_\text{AB}\bra{\psi}\otimes\rho_\text{C}
\end{equation}
for some $\rho_\text{C}$ of subsystem~C.
Thus $I(A;B|E)/2$ in Def.~(\ref{eq:squashed}) coincides with
$S_\text{vN}(\rho_\text{A})$ for any pure state~$\ket{\psi}_\text{AB}$
with reduced density matrix $\rho_\text{A}$.

Squashed entanglement is an entanglement measure based on the entropy of subsystems for bipartite pure states,
but its extension to mixed states is \emph{not} achieved through the convex-roof extension.
Instead squashed entanglement has appealing entanglement properties such as being an
entanglement monotone, being a lower bound on entanglement of formation
and being an upper bound on distillable entanglement~\cite{BDSW96,CW04}.

To obtain the monogamy relation for squashed entanglement,
consider a three-party state~$\rho_\text{ABC}$ and its extension $\rho_\text{ABCE}$.
The chain rule for quantum conditional mutual information implies that
\begin{align}
    I(A;BC|E)=I(A;B|E)+I(A;C|BE).
\label{chain}
\end{align}
Minimizing $I(A;B|E)$ and~$I(A;C|BE)$ over all possible~E and~BE, respectively,
yields $E_\text{sq}\left(\rho_\text{AB}\right)$ and~$E_\text{sq}\left(\rho_\text{AC}\right)$;
therefore,
\begin{equation}
    E_\text{sq}\left(\rho_\text{A$|$BC}\right)\geq E_\text{sq}\left(\rho_\text{AB}\right)+E_\text{sq}\left(\rho_\text{AC}\right).
\label{eq:sqmono}
\end{equation}
Thus, squashed entanglement shows the monogamy inequality of entanglement for any
tripartite state~$\rho_\text{ABC}$.
However, although squashed entanglement has beautiful properties including monogamy
and being zero if and only if the state is separable~\cite{BCY10},
the difficulty of evaluating squashed entanglement makes it problematic to use in practice.
Per definition, we need to consider all possible extensions $\rho_\text{ABE}$ of~$\rho_\text{AB}$
without even a restriction on the dimension of subsystem~E.

\section{Conclusion}
\label{sec:sum}

Beginning with a gentle introduction to entanglement and monogamy of entanglement for simple cases, we constructed mathematical monogamy relations.
Rigour has been sacrificed in favour of developing an intuitive understanding of the limits to sharing entanglement and consequent limitations to sharing classical correlations as well.
We have provided references so that the reader can follow up with technical material to master the rigour in this field.

In addition to establishing the foundations,
we have highlighted various approaches to treating monogamy of entanglement
as well as key challenges in the field.
Although monogamy of entanglement and its dual polygamy relations have been studied for more than a decade,
exciting problems remain open concerning limits to sharing multi-qudit entanglement.
These limits could have ramifications for studying entanglement sharing capabilities of quantum networks and for assessing security strengths and weaknesses of quantum networks.

At a fundamental level, our understanding of monogamy and polygamy of entanglement is predicated on choices of entanglement measures,
and development of appropriate entanglement measures is an exciting field in its own right.
Furthermore choices of entanglement measures are restricted by monogamy principles:
as discussed in this article, the monogamy principle forces an entanglement measure to be additive or superadditive
so monogamy cannot be ignored in studies of entanglement measures.

\markboth{Taylor \& Francis and I.T. Consultant}{Contemporary
Physics}
\bibliography{monogamy}
\markboth{Taylor \& Francis and I.T. Consultant}{Contemporary
Physics}

\end{document}